# Prediction of new thermodynamically stable aluminum oxides

Yue Liu[1], Artem R. Oganov[1,2,3], Shengnan Wang[1], Qiang Zhu[1], Xiao Dong[1,4], and Georg Kresse[5]

[1] Department of Geosciences, State University of New York, Stony Brook, NY 11794, USA. [2] Moscow Institution of Physics and Technology, 9 Institutskiy Lane, Dolgoprudny City, Moscow Region 141700, Russia. [3] School of Materials Science, Northwestern Polytechnical University, Xi'an 710072, China. [4] School of Physics and MOE Key Laboratory of Weak-Light Nonlinear Photonics, Nankai University, Tianjin 300071, China. [5] University of Vienna, Faculty of Physics and Center for Computational Materials Science, Sensengasse 8/12, A-1090 Wien, Austria.

*Correspondence and requests for materials should be addressed to Yue Liu (yueliu18@hotmail.com) or Artem R. Oganov (artem.oganov@sunysb.edu)*

**Recently, it has been shown that under pressure, unexpected and counterintuitive chemical compounds become stable. Laser shock experiments (A. Rode, unpublished) on alumina ($Al_2O_3$) have shown non-equilibrium decomposition of alumina with the formation of free Al and a mysterious transparent phase. Inspired by these observations, with have explored the possibility of the formation of new chemical compounds in the system Al-O. Using the variable-composition structure prediction algorithm USPEX, in addition to the well-known $Al_2O_3$, we have found two extraordinary compounds $Al_4O_7$ and $AlO_2$ to be thermodynamically stable in the pressure range 330-443 GPa and above 332 GPa, respectively. Both of these compounds at the same time contain oxide $O^{2-}$ and peroxide $O_2^{2-}$ ions, and both are insulating. Peroxo-groups are responsible for gap states, which significantly reduce the electronic band gap of both $Al_4O_7$ and $AlO_2$.**

Aluminum and oxygen are among the most abundant elements in the universe. Their only stable compound, alumina $Al_2O_3$, is widely used due to its mechanical properties (e.g. as an abrasive material) and due to its very wide band gap (for example, as an optical window material in shock-wave experiments).

$Al_2O_3$ in the corundum structure (space group $R\bar{3}c$) is an important mineral in the Earth's crust. Alumina is easily incorporated into many silicates and significantly affects their physical properties [1,2]. Several phase transitions have been theoretically predicted and experimentally confirmed to occur under pressure. It was shown that corundum transforms to the $Rh_2O_3$(II)-type structure (space group *Pbcn*) at ~80 GPa [3,4] and to the $CaIrO_3$-type phase ("post-perovskite", space group *Cmcm*) above 130 GPa [5,6]. In 2007, Umemoto et al. [7] predicted a further phase transition to a $U_2S_3$-type polymorph (space group *Pnma*) at ~370 GPa.

Given the high degree of ionicity of the Al-O bond (due to the large electronegativity difference, 1.8 on the Pauling scale), and the only possible oxidation states being +3 and -2 for Al and O, respectively, the only possible stable compound seems to be $Al_2O_3$. Of course, one can also imagine a peroxide with composition $Al_2(O_2)_3=AlO_3$, but such a compound has never been reported.

While no other stable oxides are known, there is evidence for metastable $AlO_2$, which was shown to form by an interfacial reaction in the presence of a kinetic constraint during diffusion-bonding of Pt and $\alpha$-$Al_2O_3$. Raman spectroscopy has provided strong evidence for the presence of $AlO_2$ [8], which formed after heating for 24 hours in the temperature range 1200-1400°C. $AlO_2$ is a "peroxide oxide", i.e. contains peroxide ($O_2^{2-}$), and oxide ($O^{2-}$) ions. It was not clear whether $AlO_2$ or other unusual oxides are stable at any pressure-temperature conditions.

Very recently, it has been shown that even in seemingly extremely simple systems, such as Na-Cl, totally unexpected compounds ($Na_3Cl$, $Na_2Cl$, $Na_3Cl_2$, $NaCl_3$ and $NaCl_7$) become stable under pressure – these have been predicted using evolutionary crystal structure prediction method USPEX and verified by experiments [9]. If such unusual compounds exist in the "trivial" Na-Cl system, one can expect similarly unusual compounds in nearly any other system under pressure. Here we test this hypothesis on the Al-O system, and indeed predict that $Al_4O_7$ and $AlO_2$ become thermodynamically stable under high pressure.

## Computational Methodology

To predict stable Al-O oxides and their structures, we used the evolutionary algorithm USPEX [10], [11], [12] in its variable-composition mode [13] at pressures 0, 50, 100, 150, 200, 300, 400, 500 GPa. The reliability of USPEX has been demonstrated many times before – e.g. [9], [14], [15], [16], [17], [18], [17], [18]. Modern methods have shown remarkable power to predict novel unexpected compounds – e.g. in the Na-Cl [9], Mn-B [19], Mg-C [20] and Na-Si [21] systems. Stable compositions were determined using the convex hull construction: a compound is thermodynamically stable when its enthalpy of formation from the elements and from any other compounds is negative. Enthalpy calculations and structure relaxations were done using density functional theory (DFT) within the Perdew-Burke-Ernzerhof (PBE) generalized gradient approximation (GGA) [22], as implemented in the VASP code [23]. These calculations were based on the all-electron projector-augmented wave (PAW) method [24] and plane wave basis sets with the kinetic energy cutoff of 600 eV and uniform $\Gamma$-centered k-point meshes with reciprocal-space resolution of $2\pi*0.02$ Å$^{-1}$. The first generation of structures/compositions was produced randomly with the use of space group symmetries (using algorithm [12]); the lowest-fitness 60% of the structures/compositions were allowed to produce child structures/compositions (fitness being defined as the difference between enthalpy of the structure and the convex hull). Initial structures were allowed to have up to 20 atoms in the unit cell, but this range was allowed to change in subsequent generations as a result of evolution. Child structures/compositions were created in the following manner: 20% by random symmetric generator, 40% by heredity, 20% by softmutation, and 20% by atomic transmutation. In this work, we first performed searches in the entire Al-O system with up to 20 atoms/cell, and have found only $Al_2O_3$ and oxygen-enriched phases $Al_4O_7$ and $AlO_2$. Then we did additional focused searches in a narrower compositional range $Al_2O_3$-O, and obtained the same result.

After USPEX predictions, we selected structures on the convex hull and close to it, and relaxed them carefully at pressures 0, 10, …, 520 GPa.. These calculations have confirmed stability of three oxides – well-known $Al_2O_3$ and non-classical $AlO_2$ and $Al_4O_7$. For these compounds, we also computed their electronic band structures. For accurate estimates of the band gaps, we have used the HSE hybrid functional [25]. Phonon frequencies throughout the Brillouin zone were calculated using the finite displacement approach as implemented in the Phonopy code [26,27], and these calculations confirmed that these phases are dynamically stable at pressure ranges where our enthalpy calculations predict their thermodynamic stability.

## Results

**Stable compounds in the Al-O system.** At all pressures in the range 0-500 GPa, the known compound - $Al_2O_3$ - is found to be thermodynamically stable. In agreement with previous works we find the same sequence of phase transitions – from corundum to the $Rh_2O_3$(II)-type structure at 100 GPa, then to the $CaIrO_3$-type structure at 130 GPa, and then to the $U_2S_3$-type phase at 394 GPa.

The computed thermodynamics of Al-O compounds (convex hull diagrams) are shown in Fig. 1. $Al_4O_7$ and $AlO_2$ begin to show competitive enthalpies of formation at pressures above 300 GPa and have stability fields at 330-443 GPa and at >332 GPa, respectively. From Fig. 1, one can see that at 500 GPa the enthalpy of formation of $AlO_2$ from $Al_2O_3$ and O is impressively negative, -0.12 eV/atom. The predicted pressure-composition phase diagram is shown in Fig. 2. To assess the effect of temperature, we performed quasi-harmonic free energy calculations for $AlO_2$, $Al_2O_3$ and O. We found that temperature stabilizes the formation of $AlO_2$ and expand its stability field: at T= 300 K it becomes stable at 275 GPa, at T = 2020 K at 235 GPa, and at 3140 K at 200 GPa.

**Structures of stable compounds: $Al_4O_7$ and $AlO_2$.** Structures of the stable phases of $Al_2O_3$ have been discussed elsewhere, so here we focus only on the new compounds, $Al_4O_7$ and $AlO_2$. Each of these compounds has only one stable structure up to 500 GPa, and both contain at the same time oxide $O^{2-}$ and peroxide $[O-O]^{2-}$ anions, i.e. both can be described as "oxide peroxides". At normal conditions, the O-O bond lengths are [28] 1.21 Å in the $O_2$ molecule, 1.28 Å in the superoxide $O_2^-$ ion, and 1.47 Å in the peroxide $O_2^{2-}$ ion. In $Al_4O_7$ the O-O bond length is 1.43 Å at 400 GPa, in $AlO_2$ it is 1.38 Å at 500 GPa – clearly indicating the presence of peroxide-ions.

The chemical formulas of these compounds can be obtained from $Al_2O_3$ by consecutive replacement of $O^{2-}$ by $O_2^{2-}$ (which has the same charge): taking two formula units $Al_4O_6$ and replacing $O_2 \rightarrow O_2^{2-}$, we obtain $Al_4O_5(O_2)=Al_4O_7$, and doing the same replacement again, we obtain $Al_4O_4(O_2)_2=AlO_2$. These are indeed the structural formulas: $Al_4O_5(O_2)$ for $Al_4O_7$, and $Al_4O_4(O_2)_2$ for $AlO_2$. These structures are shown in Figs. 3 and 4, and their parameters are given in Table 1.

## Discussion

**Properties of the new phases.** Phonon dispersion curves of $Al_4O_7$ and $AlO_2$, computed at 400 and 500 GPa, respectively, are shown in Fig. 5. Both phases are dynamically stable and display a continuum of phonon energies, i.e. absence of decoupled O-O vibrational modes of peroxo-groups, because at high pressure Al-O modes have frequencies comparable to O-O modes. At the same time, in the electronic structure, there are clearly defined dispersive bands of peroxo-groups, and these play an important role, as we discuss below. Both phases are dynamically and mechanically stable, as shown by their computed phonons, elastic constants, and evolutionary metadynamics [29] simulations, also enabled in the USPEX code and allowing one to explore possible phase transitions due to cell distortions and atomic displacements. We have confirmed that there are indeed no distortions or modulations that could lead to more stable structures.

All the predicted phases are insulating and show very distinct electronic structure compared with $Al_2O_3$. At 400 GPa, the computed DFT band gaps are 6.93 eV for *Pnma*-$Al_2O_3$, 2.51 eV for $Al_4O_7$, 2.92 eV for $AlO_2$. We recall that DFT calculations significantly underestimate band gaps, while hybrid functionals and GW approximation give much better band gaps, typically within 5-10% of the true values. Fig. 6 shows band gaps as a function of pressure, computed using the GGA (PBE functional), hybrid HSE functional [25] and GW approximation [30], [31]; one can see that GGA band gaps are ~30% underestimates; HSE band gaps practically coincide with the most accurate GW values for $AlO_2$, but are 0.2-1.1 eV lower for $Al_2O_3$ and $Al_4O_7$. At all levels of theory, $Al_4O_7$ and $AlO_2$ come out to have band gaps ~2 times lower than the band gap of $Al_2O_3$.

This band gap reduction for $Al_4O_7$ and $AlO_2$ originates from the additional low conduction band in the middle of the band gap. Our calculations (Fig. 7) show that these low conduction bands can be unequivocally assigned to the peroxo-groups. In both $Al_4O_7$ and $AlO_2$, both gap states - the HOMO (highest occupied molecular orbital) and LUMO (lowest unoccupied molecular orbital) - come from peroxo-groups. Together with low compressibility of the peroxo-groups (between 300 GPa and 500 GPa, the O-O distance changes from 1.37 to 1.38 Å and from 1.46 to 1.42 Å in $AlO_2$ and $Al_4O_7$, respectively), this explains why the band gaps of $Al_4O_7$ and especially $AlO_2$ are practically independent of pressure in a wide pressure range (Fig. 6). As Fig. 8 shows, projected densities of states show only small contributions from Al, thus indicating a high degree of ionicity. Indeed, Bader charges [32] are +2.44 of Al, -0.83 of O1 (peroxide anion) and -1.61 of O2 (oxide anion) in $AlO_2$ at 400 GPa.

While the band gaps computed by DFT (PBE functional) are, as expected, significantly underestimated, the energetics are accurate. We have tested this by computing the energy and enthalpy of the reaction

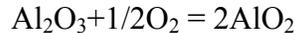
$$Al_2O_3 + 1/2 O_2 = 2AlO_2$$

using the combined exact exchange (EXX) and random phase approximation (RPA) technique [33], [34], [35]. At 300 GPa obtained the following energies (enthalpies) for this reaction: 0.1727 eV/atom (-0.0113 eV/atom) for the RPA+EXX method and 0.1782 eV/atom (0.0330 eV/atom) for PBE. At 500 GPa the results are 0.1284 eV (-0.1200 eV/atom) for RPA+EXX and 0.1371 eV (-0.1150 eV/atom) for PBE. These calculations fully confirm our findings and give additional insight:

(1) In both PBE and EXX+RPA the new compounds are stabilized by the *P*V*-term in the free energy, rather than by the internal energy. This originates from the low packing efficiency in elemental oxygen, which remains a molecular solid in the entire pressure range studied here. For this reason we can expect increased reactivity of oxygen, and stabilization of oxygen-rich compounds (such as peroxides) at high pressures.
(2) Results of the PBE and EXX+RPA are quantitatively similar, especially at 500 GPa, where the difference is only 5 meV/atom.
(3) At the EXX+RPA level of theory the new compounds predicted here are even more stable than at the PBE level.

## Conclusions

Systematic search for stable compounds in the Al-O system at pressures up to 500 GPa revealed two new stable compounds ($AlO_2$ and $Al_4O_7$); their stability fields are above 332 GPa and in the range 330-443 GPa, respectively. Our analysis reveals that insulating compounds $AlO_2$ and $Al_4O_7$ exhibit significantly ionic character, both contain peroxide $[O-O]^{2-}$ and oxide $O^{2-}$ anions and therefore belong to the exotic class of "peroxide oxides". Electronic levels of the peroxo-groups form gap states ("low conduction band") that lead to a twofold lowering of the band gap relative to $Al_2O_3$. Our preliminary results show that the formation of peroxo-ions and stabilization of peroxides under pressure occur in many oxide systems, and this phenomenon may play an important role in planetary interiors, with their high pressures and abundance of oxygen atoms.

**Figure 1. Convex hulls for the Al-O and Al$_2$O$_3$-O systems.** For the end members we used the theoretically predicted lowest-enthalpy structures from this work and Ref. [27].

**Figure 2. Pressure-composition phase diagram of the Al$_2$O$_3$-O system.**

**Figure 3: Structure of Al$_4$O$_7$ at 400 GPa.** The structure contains 7- and 8-coordinate Al atoms (coordination polyhedra are shown). Some layers of the structure contain only oxide O$^{2-}$ ions, other layers contain both oxide O$^{2-}$ and peroxide O$_2^{2-}$ ions. Peroxo-ions have two O atoms connected by a bond; the O-O bond length is 1.43 Å.

**Figure 4: Structure of AlO$_2$ at 500 GPa.** Al atoms are in the 9-fold coordination (coordination polyhedra are shown). Oxide O$^{2-}$ and peroxide O$_2^{2-}$ ions are arranged in alternating planes. Peroxo-ions have two O atoms connected by a bond; the O-O bond length is 1.38 Å.

**Figure 5: Phonon calculations for AlO$_2$ (left) and Al$_4$O$_7$ (right) at 400 GPa.**

**Figure 6: Band gaps of AlO$_2$, Al$_4$O$_7$ and Al$_2$O$_3$ as a function of pressure with (a) PBE functional, (b) HSE functional and GW method (filled symbols - HSE results, open symbols - GW results).**

**Figure 7: Band structures of AlO$_2$, Al$_4$O$_7$ and Al$_2$O$_3$ at 400 GPa.**

**Figure 8: Projected electronic densities of states of AlO$_2$ and Al$_4$O$_7$.** (a) AlO$_2$: O1 is a peroxide site and O2 is an oxide ion; (b) Al$_4$O$_7$: O2 and O3 are atoms from two different peroxide ions, and O1, O4, O5 are oxide sites.

**Table 1: Crystal structures of Al$_4$O$_7$ at 400 GPa and AlO$_2$ at 500 GPa.**

| Al$_4$O$_7$: Space group *C*2. Lattice parameters a=4.598 Å, b=9.670 Å, c=5.094 Å, β=153.5037° | | | | |
|---|---|---|---|---|
| | Wyckoff symbol | x | y | z |
| Al1 | 4c | 0.2700 | 0.0001 | 0.5216 |
| Al2 | 2a | 0 | 0.2580 | 0 |
| Al3 | 2a | 0.5000 | 0.2859 | 0 |
| O1 | 2a | 0 | 0.4399 | 0 |
| O2 | 4c | 0.2624 | 0.1299 | 0.9110 |
| O3 | 4c | 0.2534 | 0.3140 | 0.5046 |
| O4 | 2b | 0.5000 | 0.1324 | 0.5000 |
| O5 | 2a | 0.5000 | 0.4568 | 0 |
| AlO$_2$: Space group *P*2$_1$/*c*. a=4.664 Å, b=2.304 Å, c=4.726 Å, β=90.75 | | | | |
| Al | 2a | 0.2217 | 0.2750 | 0.6315 |
| O1 | 2a | 0.1339 | 0.7582 | 0.8768 |
| O2 | 2a | 0.5016 | 0.1831 | 0.3849 |

## Acknowledgments

We thank the National Science Foundation (EAR-1114313, DMR-1231586), DARPA (Grants No. W31P4Q1210008 and No. W31P4Q1310005), the Government of Russian Federation (grant No. 14.A12.31.0003), and Foreign Talents Introduction and Academic Exchange Program (No. B08040). Calculations were performed on XSEDE facilities and on the cluster of the Center for Functional Nanomaterials, Brookhaven National Laboratory, which is supported by the DOE-BES under contract No. DE-AC02-98CH10086.

## Author contribution statement

Author contributions: Y.L., Q.Z., S.N.W. and G.K. performed and analyzed calculations, Y.L., S.N.W. and A.R.O. wrote the paper, X.D. provided technical assistance with calculations.

## Additional information

**Competing financial interests:** The authors declare no competing financial interests.

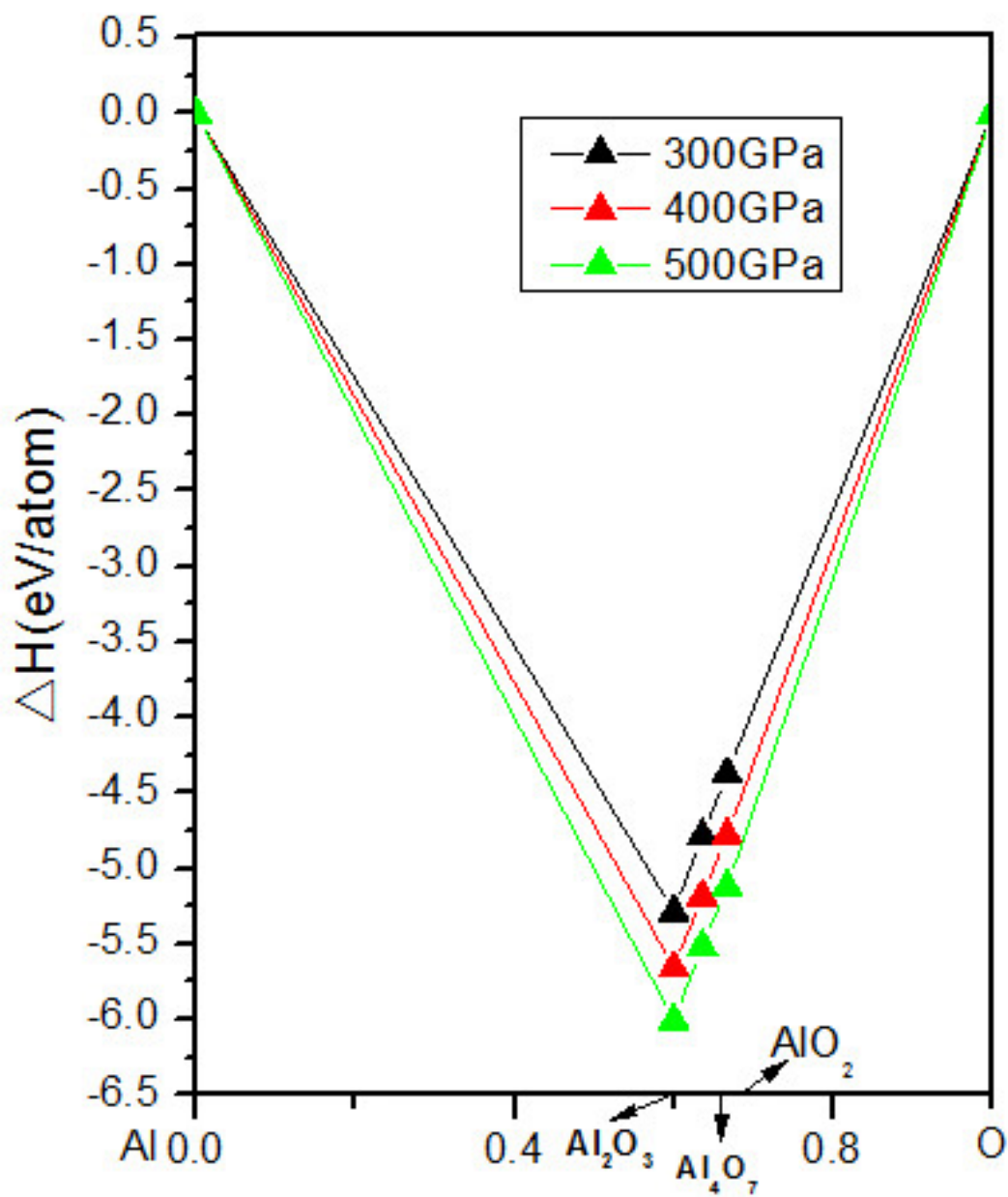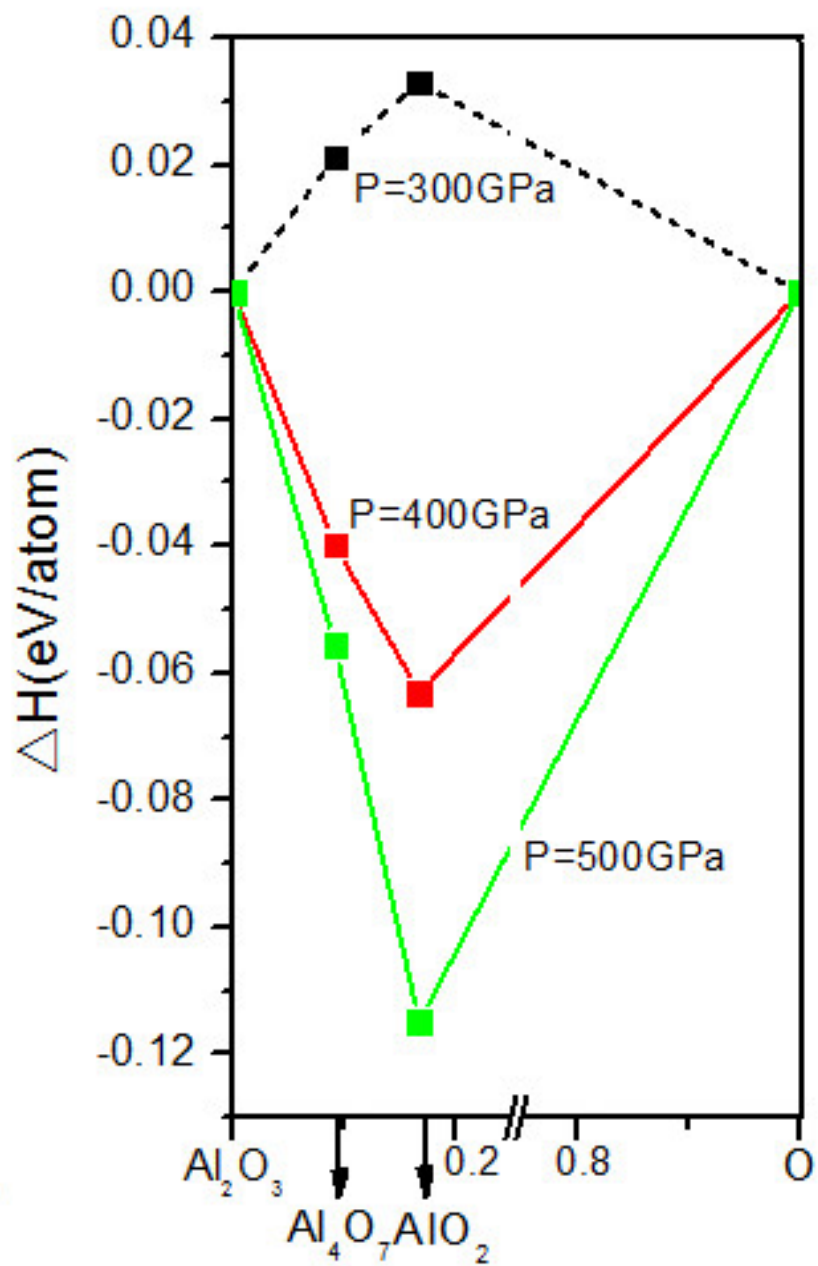

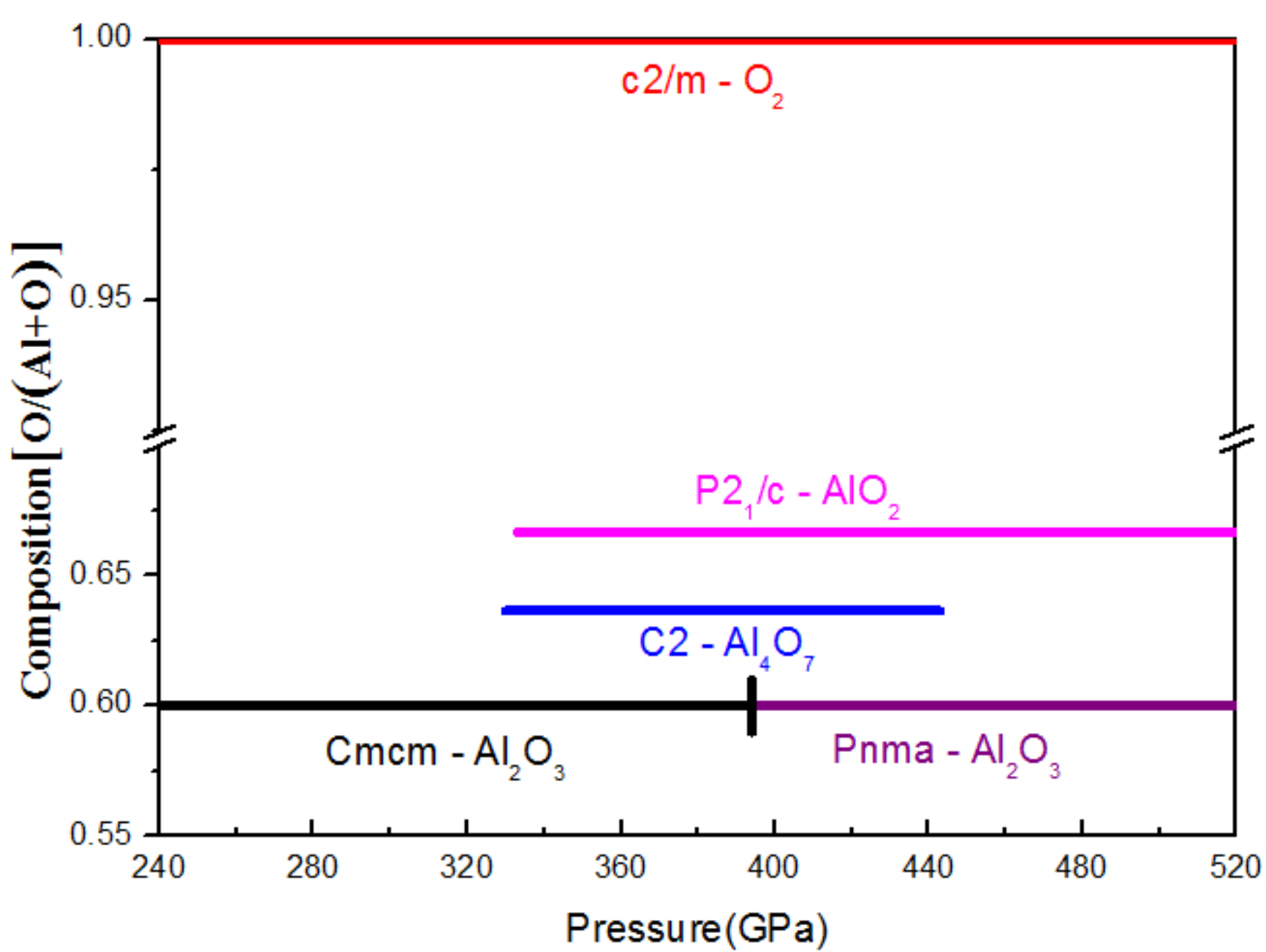

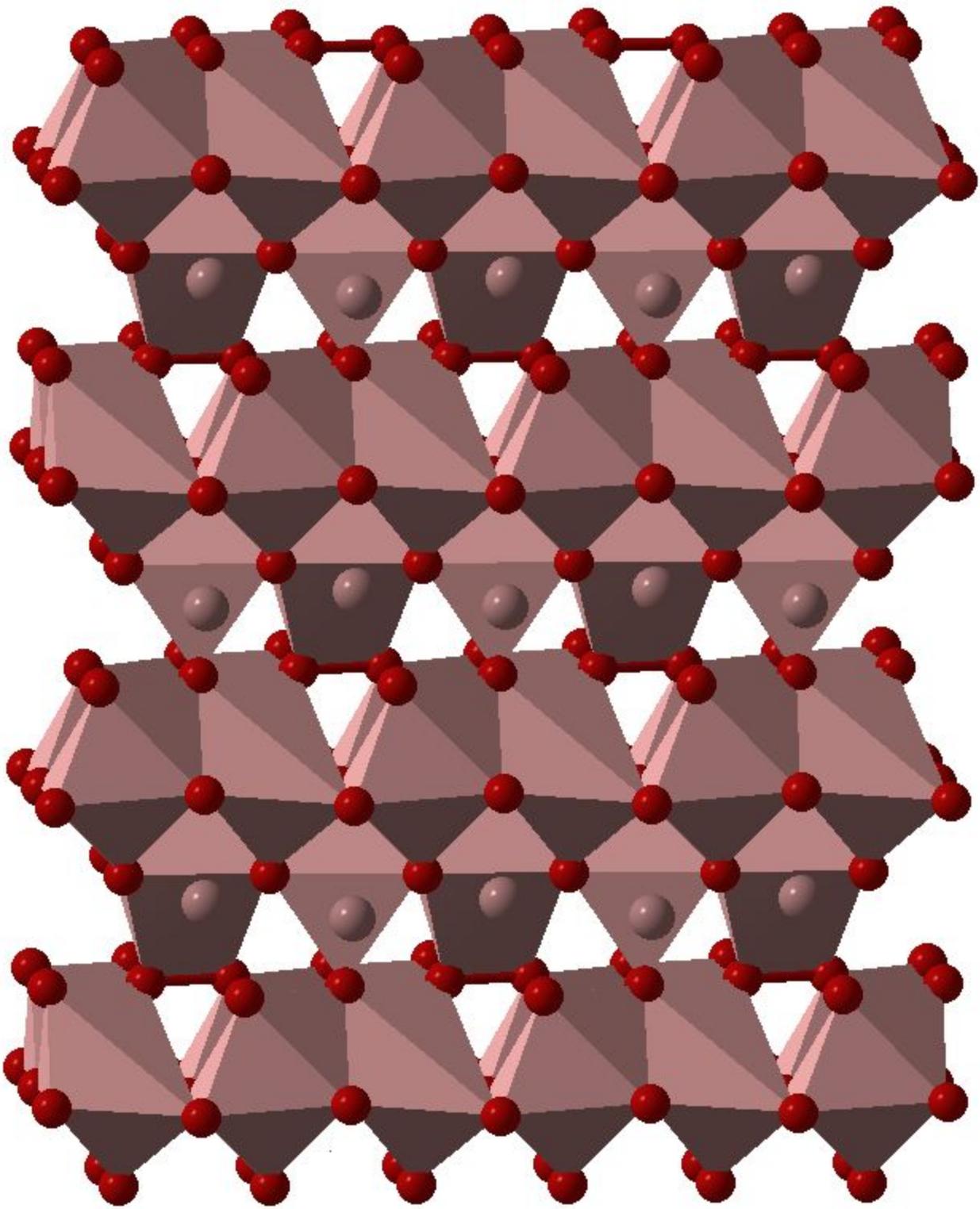

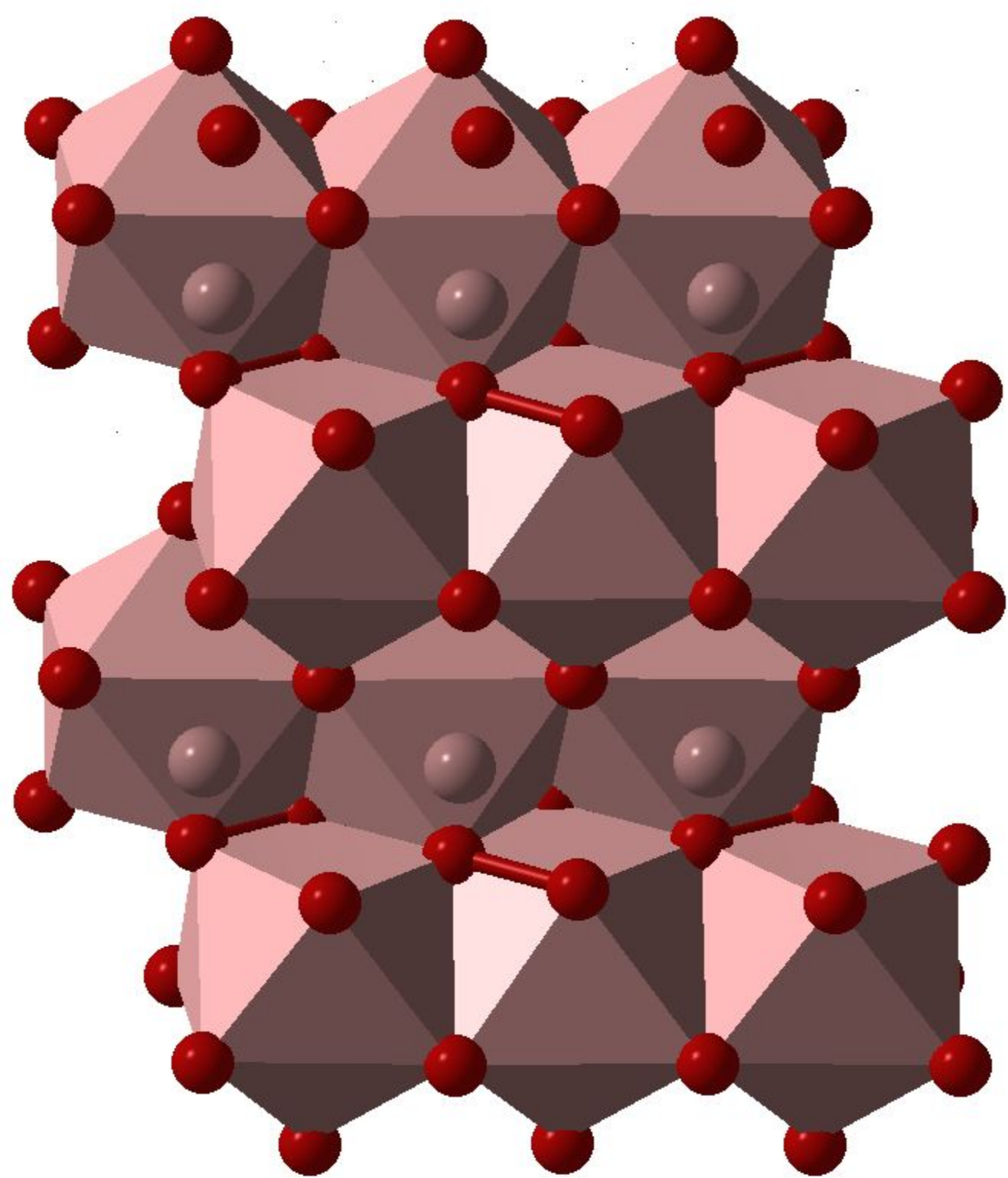

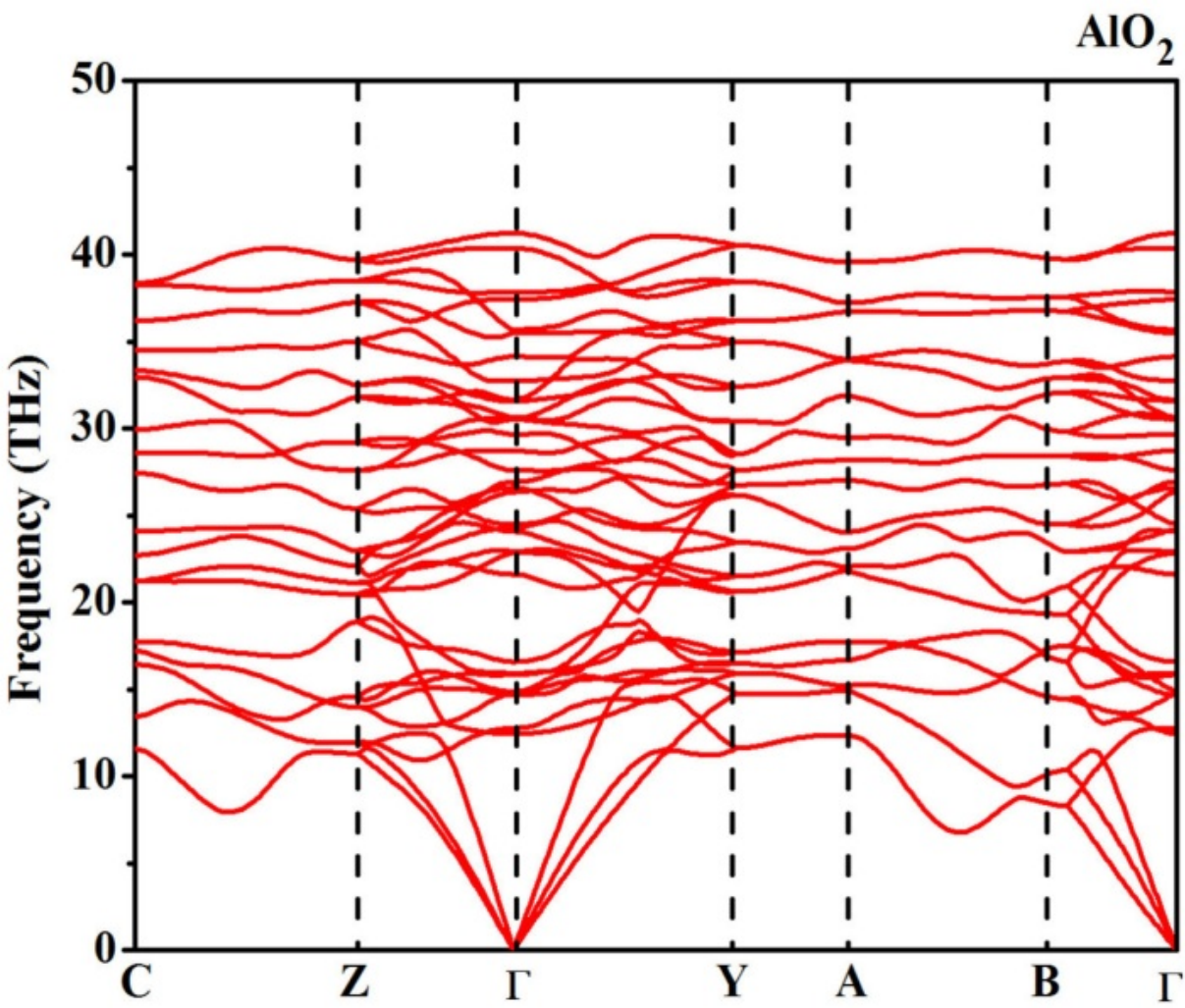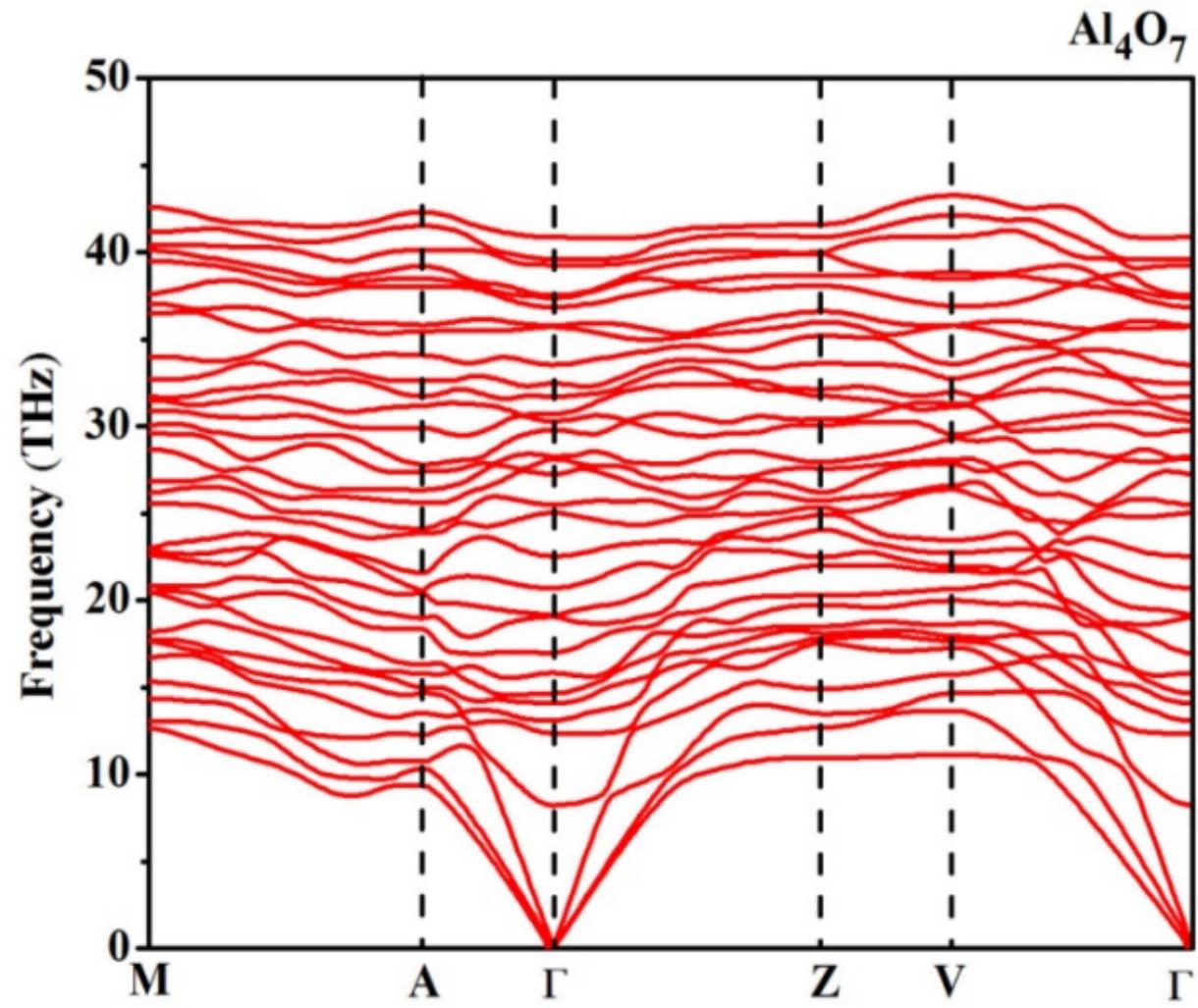

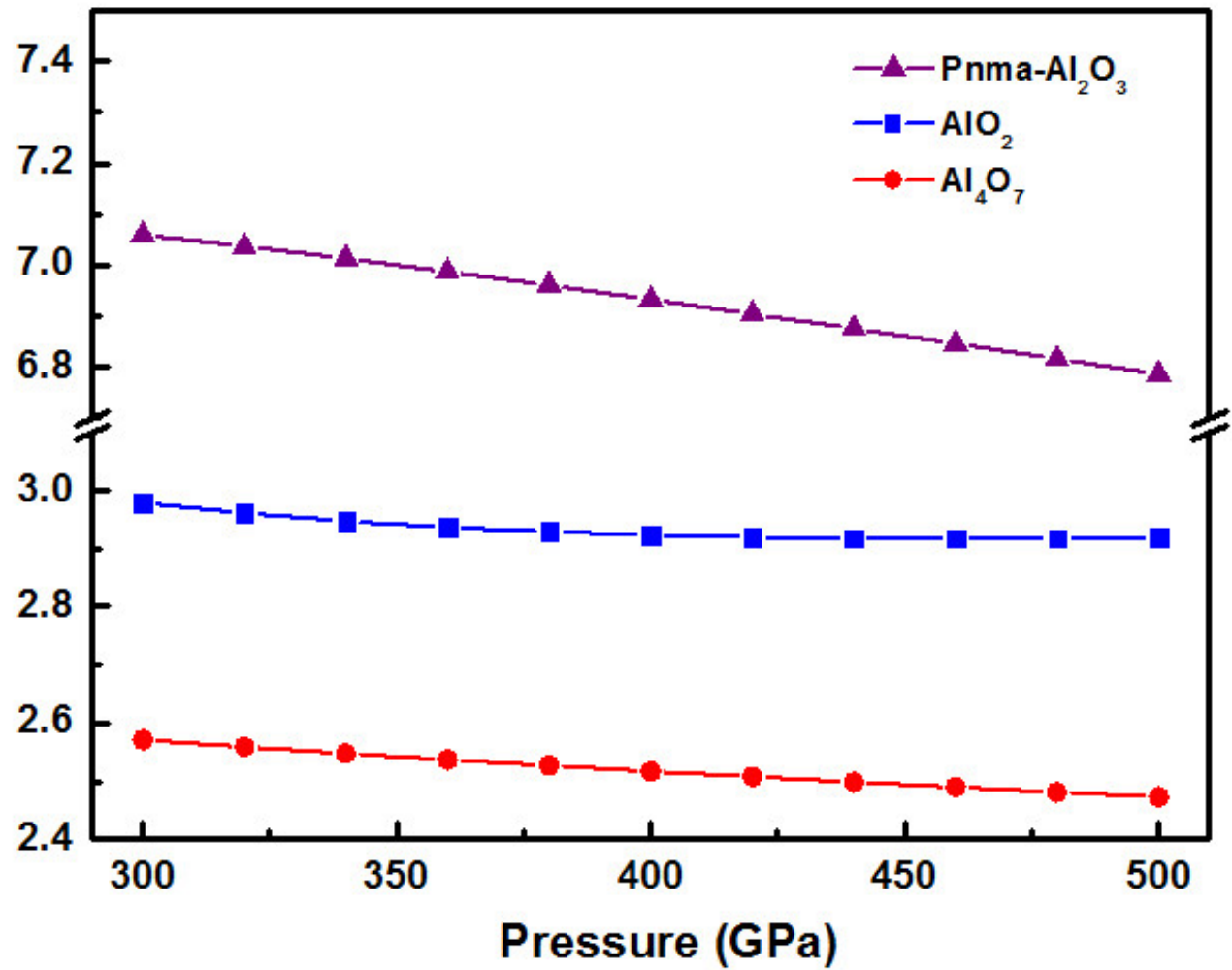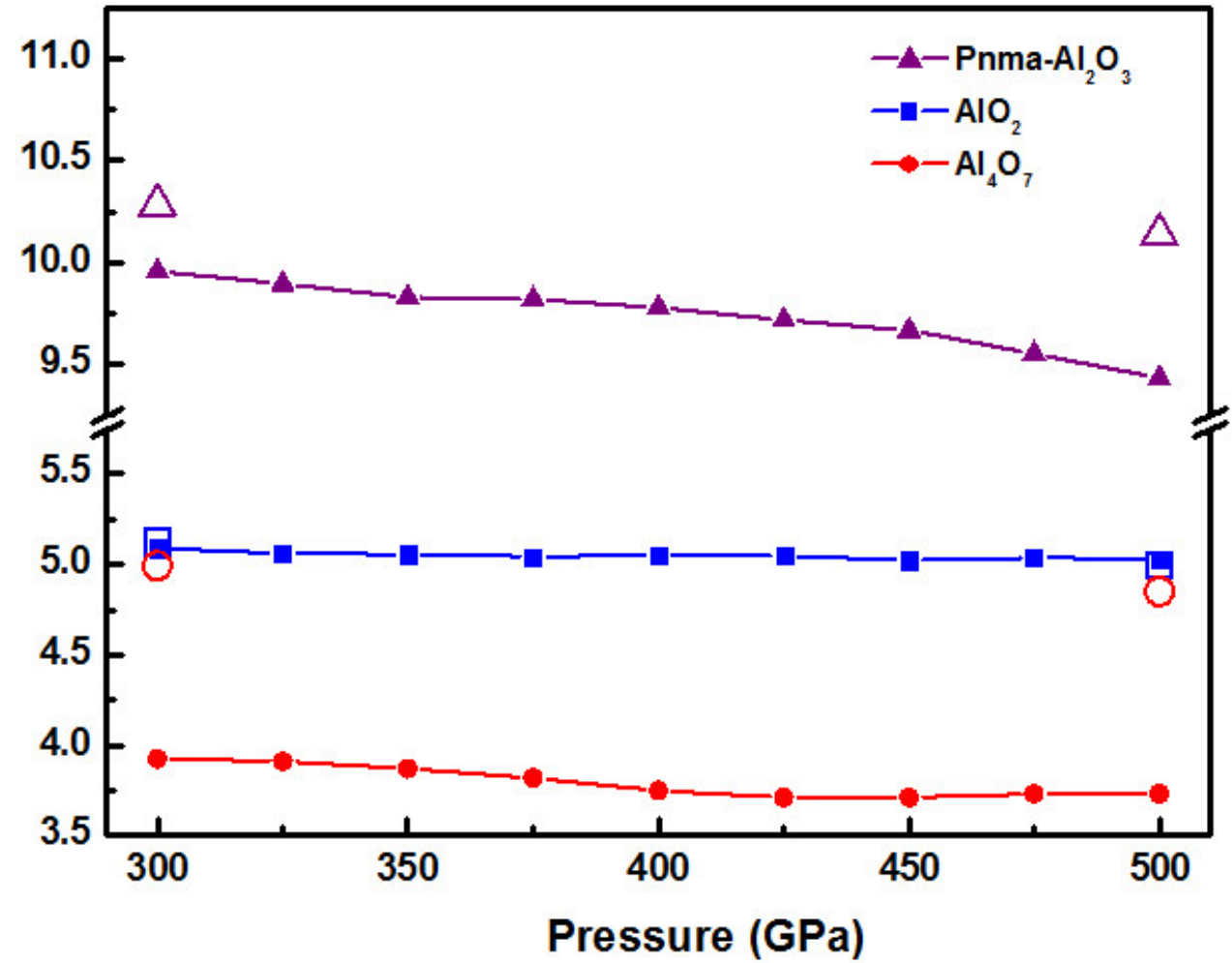

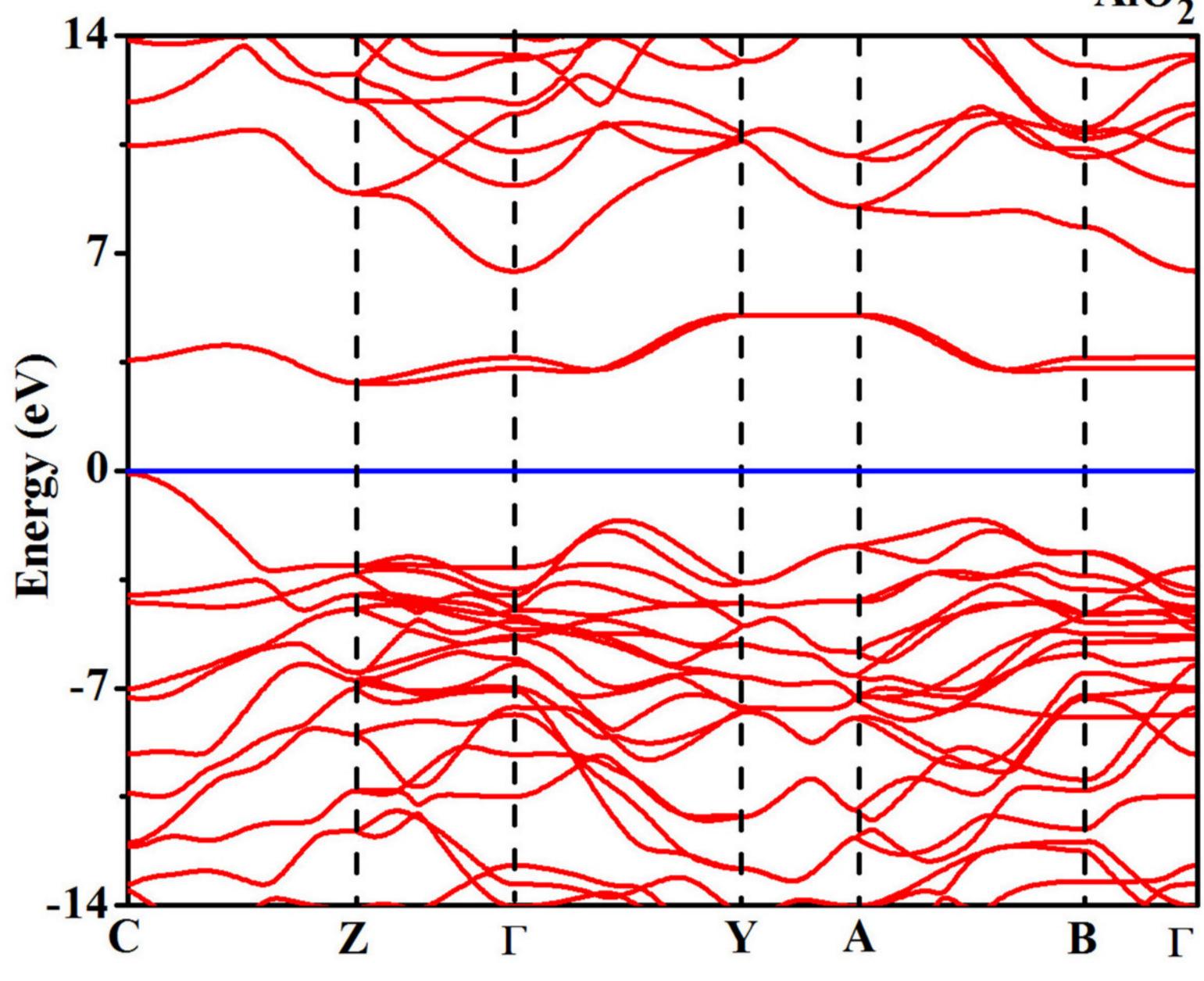
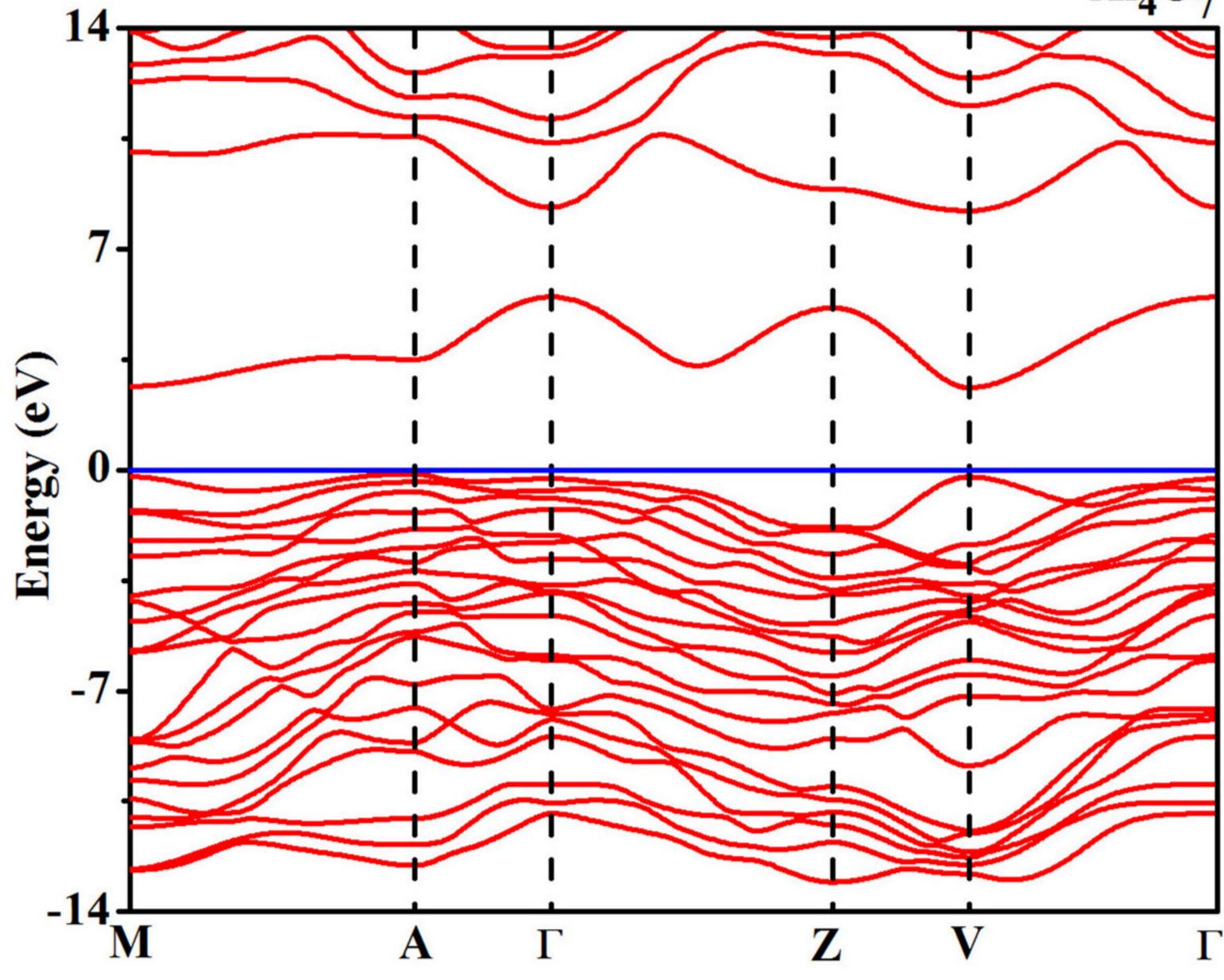
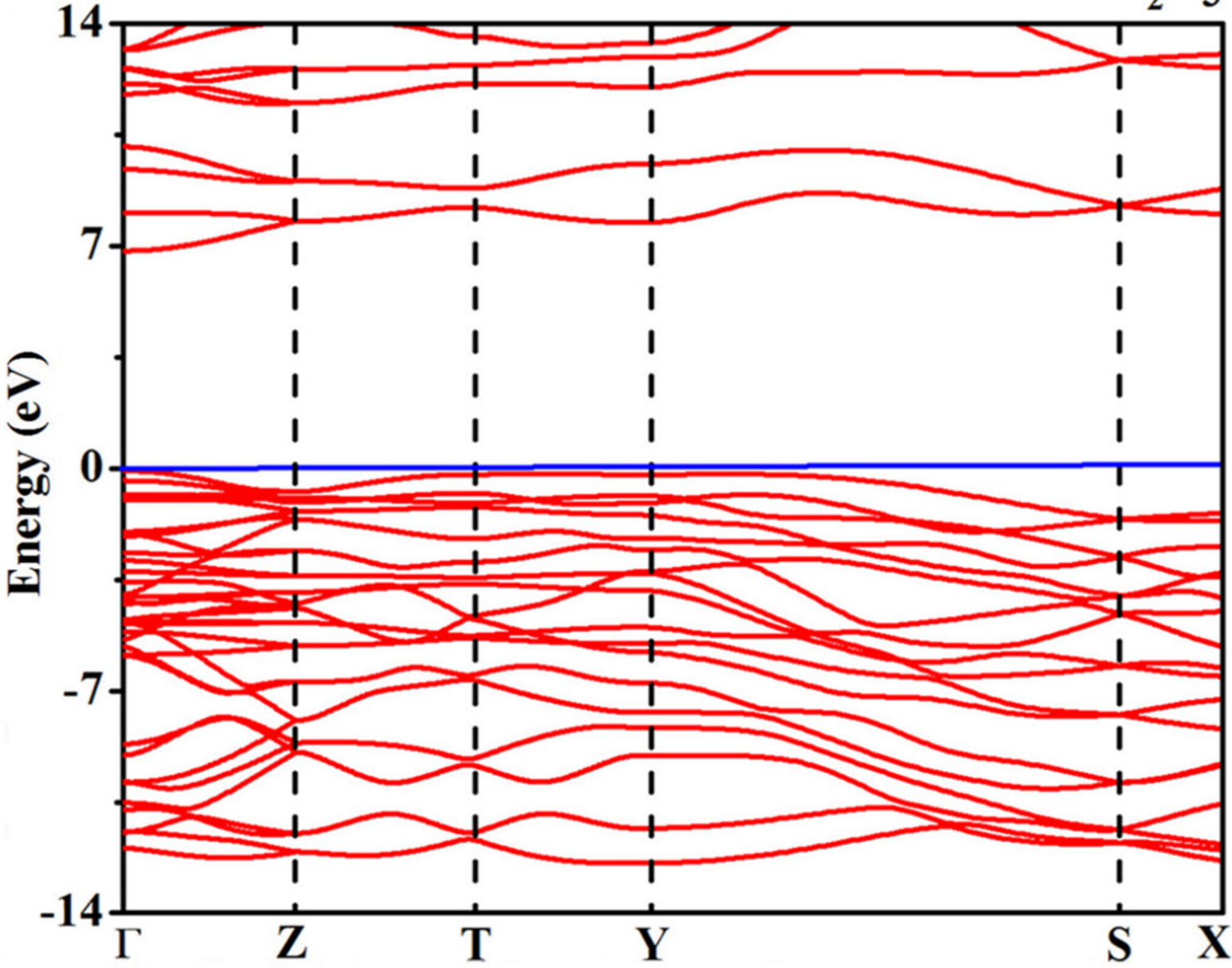

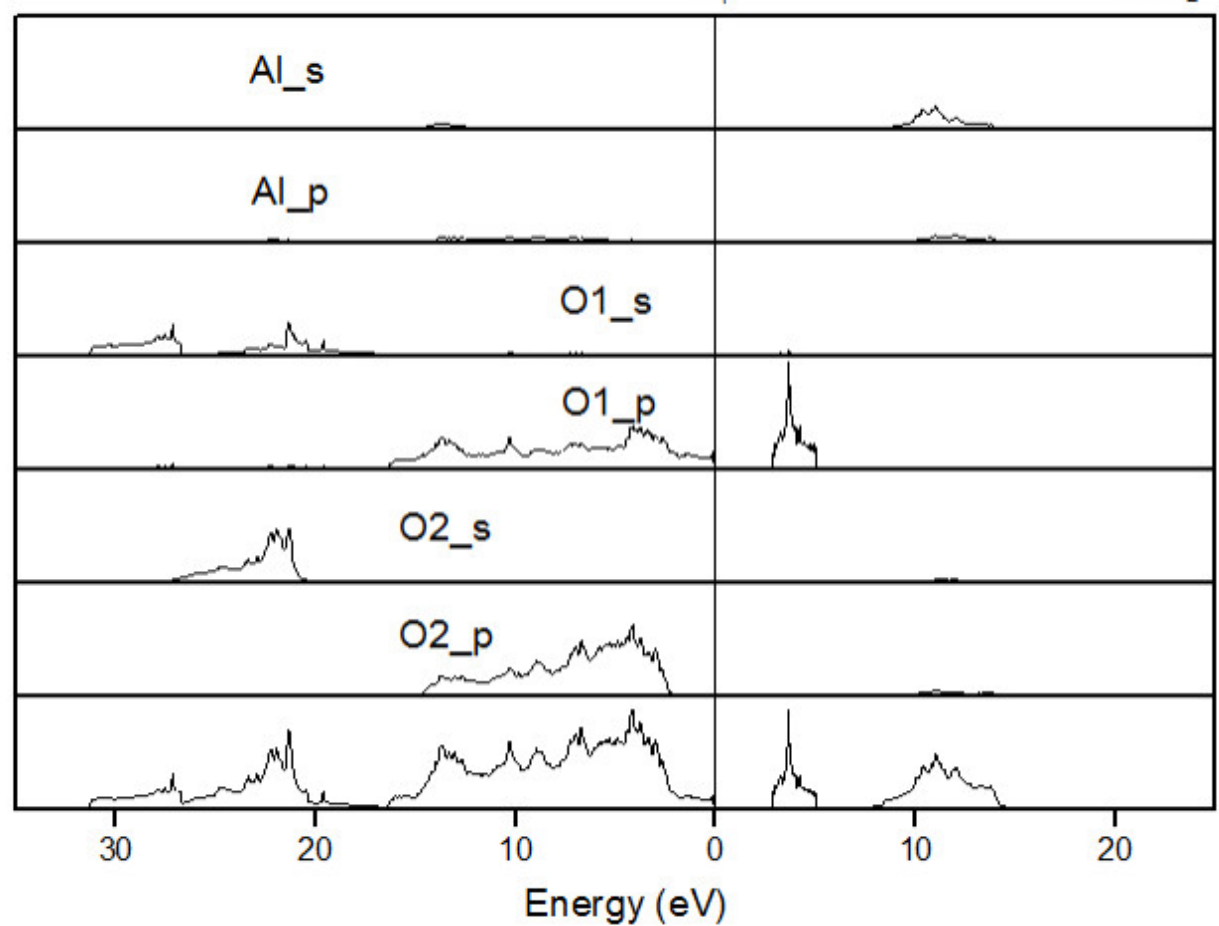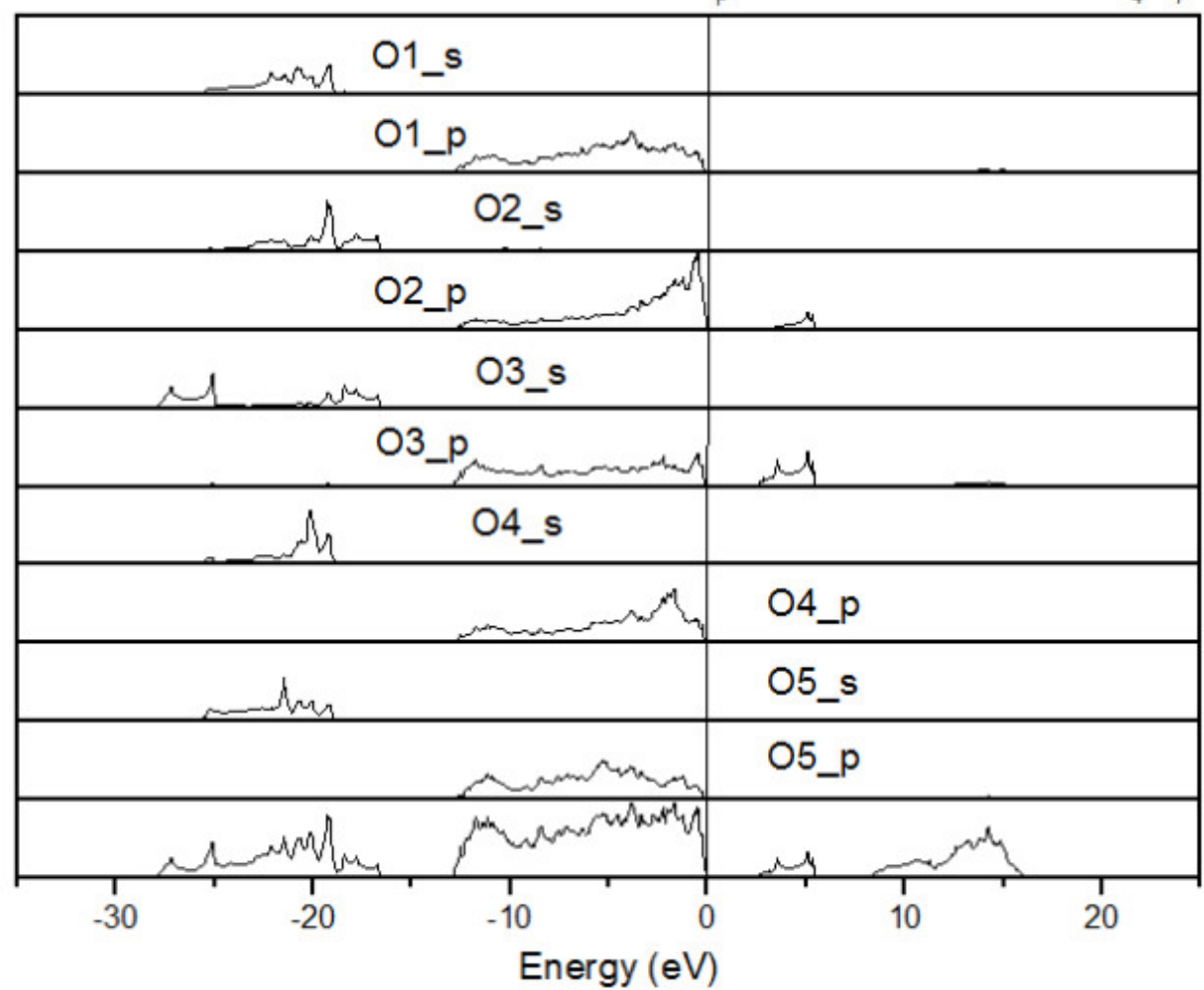